\documentclass[twocolumn]{IEEEtran}

\usepackage{amsmath,amsfonts,amsthm,amssymb}
\usepackage{mathtools}
\usepackage{nicematrix}
\usepackage{bm}
\usepackage{mathrsfs}
\usepackage{graphicx}
\usepackage{epstopdf}
\usepackage{float}
\usepackage{caption}
\usepackage{balance}
\usepackage{subfig}
\usepackage{cite}
\usepackage{multirow}
\usepackage{algorithm}
\usepackage{algpseudocode}
\usepackage{xcolor}
\def\BibTeX{{\rm B\kern-.05em{\sc i\kern-.025em b}\kern-.08em
		T\kern-.1667em\lower.7ex\hbox{E}\kern-.125emX}}

\allowdisplaybreaks[4]

\newcommand{\rev}{\textcolor{black}}

\begin{document}

\title{\rev{Delay-Doppler Diversity Analysis of ODDM Modulation in Satellite Communication Systems}}
\author{Yu Liu, Cunhua Pan, Tantao Gong, Yinlu Wang, Ming Chen
\thanks{\textit{(Corresponding author: Ming Chen.)}}
\thanks{Yu Liu, Cunhua Pan, Tantao Gong, Ming Chen are with are with the National Mobile Communications Research Laboratory, Southeast University, Nanjing, 210096, China 
(e-mail: \{liuyu\_1994, cpan, gongtantao, chenming \}@seu.edu.cn).}
\thanks{Yinlu Wang is with  the School of Communication and Information Engineering, Nanjing University of Posts and Telecommunications, Nanjing, 210003, China (e-mail: yinluwang@njupt.edu.cn).}
}
\maketitle

\begin{abstract}
	\rev{
		Orthogonal delay-Doppler division multiplexing (ODDM) modulation offers a promising solution to the problem of severe Doppler effects in low earth orbit (LEO) satellites communications.
		It has been suggested in the recent literature that ODDM modulation can extract full delay-Doppler (DD) diversity, yet a rigorous analysis has not been presented.
		In this paper, we present a formal analysis of the DD diversity achieved by ODDM modulation along with supporting simulations.
		Specifically, the analysis and simulations reveal that the asymptotic DD diversity order of ODDM modulation is one, and this order is achieved at lower bit error rate (BER) values for increased frame sizes.
		We also present low-complexity detector for ODDM modulation based on the Orthogonal approximate message passing (OAMP) algorithm, and show that this detector extracts full DD diversity with the reduced complexity.
	}
\end{abstract}


\section{Introduction}
In order to achieve the goal of global seamless communications, the low-earth-orbit (LEO) satellite communication is incorporated for the next generation mobile network. However, the high mobility of LEO satellites brings the severe Doppler effects, which is undoubtedly harmful to communication reliability.
Recently, a series of delay-Doppler (DD) domain modulation has attracted significant research attention, which aims to convert the doubly-selective channel in the time-frequency (TF) domain into the almost non-fading channel in the DD domain. 
As a representative work in the DD domain modulation, the authors of \cite{2017HadaniOTFS} proposed the inverse symplectic finite Fourier transform (ISFFT) based orthogonal time frequency space (OTFS) modulation. However, this ISFT-spread OTFS scheme employs TF domain orthogonal pulses that do not match the DD domain resolution, resulting in complicated input-output relation and performance degradation \cite{2024SwaroopZak-OTFS}.
Henceforth, the authors of \cite{2022HaiODDM} proposed orthogonal delay-Doppler division multiplexing (ODDM) modulation as a promising and practical DD domain multi-carrier modulation. 
Since orthogonal pulses match DD domain resolutions, ODDM is promising to have a simple input-output relation and practical implementation. 

%

\rev{
	ODDM modulation has achieved significantly better error performance compared to orthogonal frequency division multiplexing (OFDM) under severe Doppler conditions \cite{2022HaiODDM}. Additionally, ODDM modulation has also demonstrated comparable error performance to OTFS modulation, employing low complexity symbol detection techniques. Authors in \cite{2023ChengCLN} presented a message passing algorithm (MPA) based detector, and authors in \cite{2024TongODDM} presented a symbol detector based on conjugate gradients. 
	Owing to the similarity between ODDM and OTFS modulation, an orthogonal approximate message passing (OAMP) based detector originally developed for OTFS in \cite{2022YangOAMPOTFS} can be adapted to ODDM. This approach holds promise for exploiting the sparsity of the effective DD domain channel and achieving full DD domain diversity with reduced complexity \cite{2017OAMP,2022YangOAMPOTFS}.
}


\rev{
	The theoretical analysis on transmission performance of ODDM has not been thoroughly studied as that of OTFS yet.
	We note that the authors in \cite{2022HaiODDM} suggested that ODDM can achieve the full DD diversity. However, this suggestion has not been established through analysis or simulations. 
	To fill this gap, we aim to investigate the diversity order achieved by ODDM in LEO satellite communications with theoretical analysis and supporting simulations.
	The key findings and contributions in this work can be summarized as follows:
	\begin{enumerate}
		\item We derive the unconditional bound on bit error rate (BER) for ODDM with maximum likelihood (ML) detection by leveraging the conditional pairwise error probability (PEP). We prove that the asymptotic DD diversity order of ODDM can be one, and it is observed the diversity one region starts at lower BER values for increased frame sizes. 
		\item In attempt to extract the full diversity, we present an OAMP based detector for ODDM. The simulations indicate that this detector extracts the full diversity in the DD domain with reduced complexity. Moreover, the simulations demonstrate that ODDM achieves comparable BER performance to OTFS with significantly lower out-of-band emission (OOBE).
	\end{enumerate}
}

\section{System Model}
In this paper, we consider a point-to-point LEO satellite communication system that employs ODDM. 
Without loss of generality, let us consider ODDM transmitter as follows.
Let $M$ be the number of delay bins/subcarriers and $N$ be the number of Doppler bins/time slots, respectively. In the term of time-frequency domain, the corresponding subcarrier spacing and time slot duration are respectively given by $\Delta f$ and $T$, which obey $\Delta f=\frac{1}{T}$. In the term of delay-Doppler domain, the delay resolution of the signal is $\frac{1}{M\Delta f}$, and the Doppler resolution of the signal is $\frac{1}{NT}$.
\rev{Let $\mathbf{X} \in \mathbb{A}^{M\times N}$ be the transmitted symbol matrix in the DD domain, where $\mathbb{A}$ is the constellation set.}
To be specific, let $x[m,n]$ represent the $(m,n)$-th element in $\mathbf{X}$, and $\mathbf{x}\in\mathbb{A}^{MN\times1}$ be the vectorization of $\mathbf{X}$.
According to \cite{2022HaiODDM}, ODDM is a multicarrier (MC) modulation, which aims to modulate DD domain symbols into $M$ staggered MC symbols, with symbol period of $NT$, subcarrier spacing of $\frac{1}{NT}$ and symbol interval of $\frac{T}{M}$, respectively. Then the CP-free waveform of the DD plane MC modulation is given by
\begin{equation}
	s(t)=\sum_{m=0}^{M-1}\sum_{n=0}^{N-1}x[m,n]\hat{g}_\text{tx}\left(t-m\frac{T}{M}\right)e^{j2\pi\frac{n}{NT}\left(t-m\frac{T}{M}\right)},
\end{equation}
where $\hat{g}_\text{tx}(t)$ is the transmit pulse. Note that $\hat{g}_\text{tx}(t)$ is a DD domain orthogonal pulse with respect to the DD plane resolutions. 
It can be obtained by $\hat{g}_\text{tx}(t)=\sum_{\dot{n}=0}^{N-1}a(t-\dot{n}T)$, 
where $a(t)$ denotes a time-symmetric real-valued square-root Nyquist pulse that spans a time duration of $2Q\frac{T}{M}$, with $Q$ being an integer and $2Q\ll M$. The pulse satisfies $a(t)=0$ for $t\notin \left(-Q\frac{T}{M},Q\frac{T}{M}\right)$.
Without loss of generality, suppose that $\int_{-\infty}^{+\infty}|a(t)|^2\mathrm{d}t=\frac{1}{N}$, thus $\int_{-\infty}^{+\infty}|\hat{g}_\text{tx}(t)|^2\mathrm{d}t=1$.
Finally, considering the cyclic prefix (CP), we extend the definition of $\hat{g}_\text{tx}$ to $u_\text{cp}(t)=\sum_{\dot{n}=-1}^{N-1}a(t-\dot{n}T)$.
Without loss of generality, we assume that the maximum delay and Doppler shift of the channel are $(L-1)\frac{T}{M}$ and $K\frac{1}{NT}$, respectively.
Then, the CP-included ODDM waveform spanning over $-(L+Q-1)\frac{T}{M}\le t \le NT+(Q-1)\frac{T}{M}$ becomes
\begin{equation}
	s_\text{cp}(t)=\sum_{m=0}^{M-1}\sum_{n=0}^{N-1}x[m,n]u_\text{cp}\left(t-m\frac{T}{M}\right)e^{j2\pi \frac{n}{NT}\left(t-m\frac{T}{M}\right)},
\end{equation}
where $u_\text{cp}(t)=u(t)$ for $t\in (-Q\frac{T}{M},(N-1)T+Q\frac{T}{M}$ and $u_\text{CP}(t)=0$ for $t\in(-T+Q\frac{T}{M},-Q\frac{T}{M})$.

At the receiver side, the received ODDM signals for $-(L+Q-1)\frac{T}{M} \le t \le NT+(L+Q-2)\frac{T}{M}$ can be expressed as 
\begin{equation}
	\begin{aligned}
		y(t)&=\sum_{p=1}^{P}h_p s_\text{cp}(t-\tau_p)e^{j2\pi\nu_p(t-\tau_p)}+z(t) \\
		&=\sum_{p=1}^{P}\sum_{m=0}^{M-1}\sum_{n=0}^{N-1}h_px[m,n]u_\text{cp}\left(t-(m+l_p)\frac{T}{M}\right) \\
		& \quad \cdot e^{j2\pi\frac{n+k_p}{NT}\left(t-(m+l_p)\frac{T}{M}\right)}e^{j2\pi\frac{k_p m}{MN}} + z(t),
	\end{aligned}
\end{equation}
where $P$ is the number of multipaths and $h_p$ refers to the channel gain of the $p$-th path. Note that, $l_p=\frac{\tau_p}{M\Delta f}$ denotes the delay index, and $\tau_p$ is the delay shift of the $p$-th path. $k_p=\frac{\nu_p}{NT}$ depicts Doppler index, and $\nu_p$ represents Doppler shift of the $p$-th path. $z(t)$ indicates the noise in the DD domain. 
Passing the matched filter, the received ODDM signal in the DD domain is obtained as \eqref{ODDM signal},
where $z[m,n]$ indicates the DD domain noise sample.
\begin{figure*}[t]
	\vspace*{-25pt}
	\begin{equation}
		y[m,n]=\begin{cases}
			\sum_{p=1}^P h_p x\left[m-l_p,[n-k_p]_N\right]e^{j2\pi\frac{k_p(l-l_p)}{MN}}+z[m,n], &\; m-l_p\ge 0 \\
			\sum_{p=1}^P h_p e^{-j2\pi\frac{[n-k_p]_N}{N}}x\left[[m-l_p]_M,[n-k_p]_N\right]e^{j2\pi\frac{k_p(l-l_p)}{MN}}+z[m,n], &\; m-l_p < 0
		\end{cases}
		\label{ODDM signal}
	\end{equation}
	\rule{\linewidth}{.5pt}
	\vspace*{-25pt}
\end{figure*}
The symbol-wise input-output relation can be transformed into a vectorized form,
\begin{equation}
	\mathbf{y}=\mathbf{Hx+z},
	\label{sparse linear problem}
\end{equation}
where $\mathbf{y}\in\mathbb{C}^{MN\times 1}$ and $\mathbf{z}\in\mathbb{C}^{MN\times 1}$ stand for the vectorized form of the received signal and noise in the DD domain. $\mathbf{H}\in\mathbb{C}^{MN\times MN}$ is the effective channel matrix, given by
\begin{equation}
	\mathbf{H} =
	\left[ \begin{matrix}
		\mathbf{H}_{0}^{0}&		&		&		\mathbf{H}_{0}^{L-1}\mathbf{D}&		\cdots&		\mathbf{H}_{1}^{0}\mathbf{D}\\
		\vdots&		\ddots&		&		&		\ddots&		\vdots\\
		\mathbf{H}_{L-1}^{L-1}&		\ddots&		\mathbf{H}_{0}^{L-1}&		\huge\textbf{0} &		&		\mathbf{H}_{L-1}^{L-2}\mathbf{D}\\
		&		\ddots&		\ddots&		\ddots&		\ddots&		\\
		\huge\textbf{0} &		\mathbf{H}_{L-1}^{M-1}&		\ddots&		\ddots&		\ddots&		\mathbf{H}_{0}^{M-1}\\
	\end{matrix} \right],
	\label{H matrix}
\end{equation}
where $\mathbf{H}_{l}^{m}=\sum_{\hat{k}=-K}^K{g\left( \hat{k}+K+1,l \right) e^{j2\pi \frac{\hat{k}(m-l)}{MN}}\mathbf{C}^{\hat{k}}}$ denotes the sub-channel block and $\mathbf{D}=\text{diag}\{1,e^{-j\frac{2\pi}{N}}\,\cdots,e^{-j\frac{2\pi(N-1)}{N}}\}$ denotes the additional phase rotation. Note that in $\mathbf{H}_l^m$, $\mathbf{C}$ is the $N\times N$ cyclic permutation matrix and $g(\cdot)$ is given by
\begin{equation}
	g(\hat{k}+K+1,l)=\begin{cases}
		h_p, \; &\text{if } \hat{k}+K+1=k_p, l=l_p, \\
		0, & \text{Otherwise}.
	\end{cases}
\end{equation}
Apparently, $\mathbf{H}$ is a sparse matrix.

\section{\rev{Diversity Analysis of ODDM}}
In this section, in order to study the diversity of ODDM, we consider an alternative formulation of input-output relation of this ODDM scheme given by
\begin{equation}
	\mathbf{y}=\bm{\Phi}(\mathbf{x})\mathbf{h}+\mathbf{z},
\end{equation}
where $\mathbf{h}=\left[h_1,h_2\cdots,h_p\right]^\text{T} \in \mathbb{C}^{P\times 1}$ is the vector of complex channel gains, and the small-scale channel coefficient $h_p$ follows $\mathcal{N}(0,1/P)$.
$\bm{\Phi}(\mathbf{x}) \in \mathbb{C}^{MN\times P}$ is the concatenated matrix given by
\begin{equation}
	\bm{\Phi}(\mathbf{x})=\left[\bm{\Xi}_1\mathbf{x}|\cdots|\bm{\Xi}_P\mathbf{x}\right],
\end{equation}
and $\bm{\Xi}_p \in \mathbb{C}^{MN\times MN}$ is defined in \eqref{Xi definition}.

For convenience, we normalize the elements of $\mathbf{x}$, thus the energy allocated to each symbol is one. The transmitting signal-to-noise ratio (SNR), denoted by $\gamma$, is therefore given by $\gamma=1/N_0$. Assuming perfect channel estimation, the conditional PEP of the transmitting symbols $\mathbf{x}$ and the receiving symbols $\hat{\mathbf{x}}$ is given by \cite{2005TseFundamentals},
\begin{equation}
	P\left(\mathbf{x}\rightarrow\hat{\mathbf{x}}|\mathbf{h},\mathbf{x}\right)=\mathcal{Q}\left(\sqrt{\frac{\|[\bm{\Phi}(\mathbf{x})-\bm{\Phi}(\hat{\mathbf{x}})]\mathbf{h}\|^2}{2N_0}}\right),
\end{equation}
where $\mathcal{Q}(\cdot)$ is the tail distribution function of the standard normal distribution. 
Henceforth, the expected PEP become
\begin{equation}
	P(\mathbf{x}\rightarrow\hat{\mathbf{x}})=\mathbb{E}_{\mathbf{h}}\left[\mathcal{Q}\left(\sqrt{\frac{\gamma\|[\bm{\Phi}(\mathbf{x})-\bm{\Phi}(\hat{\mathbf{x}})]\mathbf{h}\|^2}{2}}\right)\right].
	\label{expected PEP}
\end{equation}
Rewrite $\|\left[\bm{\Phi}(\mathbf{x})-\bm{\Phi}(\hat{\mathbf{x}})\right]\mathbf{h}\|^2$ as
\begin{equation}
	\|\left[\bm{\Phi}(\mathbf{x})-\bm{\Phi}(\hat{\mathbf{x}})\right]\mathbf{h}\|^2=\mathbf{h}^\text{H}\left[\bm{\Phi}(\mathbf{x})-\bm{\Phi}(\hat{\mathbf{x}})\right]^\text{H}\left[\bm{\Phi}(\mathbf{x})-\bm{\Phi}(\hat{\mathbf{x}})\right]\mathbf{h}.
	\label{norm expansion}
\end{equation}
The matrix $\left[\bm{\Phi}(\mathbf{x})-\bm{\Phi}(\hat{\mathbf{x}})\right]^\text{H}\left[\bm{\Phi}(\mathbf{x})-\bm{\Phi}(\hat{\mathbf{x}})\right]$ is Hermitian matrix that can be transformed into 
\begin{equation}
	\left[\bm{\Phi}(\mathbf{x})-\bm{\Phi}(\hat{\mathbf{x}})\right]^\text{H}\left[\bm{\Phi}(\mathbf{x})-\bm{\Phi}(\hat{\mathbf{x}})\right]=\mathbf{U}\bm{\Lambda}\mathbf{U}^\text{H},
\end{equation} 
\rev{where $\mathbf{U}\in\mathbb{C}^{P\times P}$ is the unitary matrix and $\bm{\Lambda}=\text{diag}\{\lambda_1^2,\lambda_2^2,\cdots,\lambda_P^2\}$ with $\lambda_p$ being the $p$-th singular value of the difference matrix $\bm{\Delta}(\mathbf{x},\hat{\mathbf{x}})=\bm{\Phi}(\mathbf{x})-\bm{\Phi}(\hat{\mathbf{x}})$.} \rev{Defining $\tilde{h}^\text{H}=\mathbf{U}^\text{H}\mathbf{h}^\text{H}$, \eqref{norm expansion} can be simplified as
	\begin{equation}
		\|\left[\bm{\Phi}(\mathbf{x})-\bm{\Phi}(\hat{\mathbf{x}})\right]\mathbf{h}\|^2=\tilde{\mathbf{h}}\mathbf{\Lambda}\tilde{\mathbf{h}}^\text{H}=\sum_{p=1}^{r}\lambda_p^2|\tilde{h}_p|^2,
		\label{simplified norm}
	\end{equation}
	where $r$ denotes the rank of the difference matrix $\bm{\Delta}(\mathbf{x},\hat{\mathbf{x}})$. }
Substituting \eqref{simplified norm} into \eqref{expected PEP}, the expected PEP can become 
\begin{equation}
	P(\mathbf{x}\rightarrow\hat{\mathbf{x}})=\mathbb{E}_{\mathbf{h}}\left[\mathcal{Q}\left(\sqrt{\frac{\gamma\sum_{p=1}^r\lambda_p^2|\tilde{h}_p|^2}{2}}\right)\right].
\end{equation}
\rev{
	According to \cite{2005TseFundamentals}, at high SNRs, the expected PEP can be simplified to get the following upper bound,}
\begin{equation}
	P(\mathbf{x}\rightarrow\hat{\mathbf{x}}) \le \prod_{p=1}^{r}\frac{1}{1+\frac{\gamma\lambda_p^2}{4P}}.
	\label{PEP 1}
\end{equation}
In the case of Rayleigh fading, \eqref{PEP 1} can be further simplified as 
\begin{equation}
	P(\mathbf{x}\rightarrow\hat{\mathbf{x}}) \le \frac{1}{\prod_{p=1}^r \lambda_p^2/P}\left(\frac{\gamma}{4}\right)^{-r}.
	\label{PEP 2}
\end{equation}
Henceforth, the upper bound on the average bit error probability (BEP) $P_\text{e}$ can be expressed as
\begin{equation}
	P_\text{e} \le \frac{1}{Q^{MN}}\sum_{\mathbf{x}\in\mathbb{A}^{MN}}\sum_{\mathbf{x}\neq\hat{\mathbf{x}}}\frac{\text{d}_\text{b}(\mathbf{x},\hat{\mathbf{x}})}{MN\log_2Q}
	\left(\prod_{p=1}^r\lambda_p^2\right)^{-1}\left(\frac{\gamma}{4P}\right)^{-r}.
	\label{BEP}
\end{equation}
where $Q$ is the modulation order and $\text{d}_\text{b}(\mathbf{x},\hat{\mathbf{x}})$ is the difference in number of information bits between $\mathbf{x}$ and $\hat{\mathbf{x}}$. From \eqref{PEP 1} and \eqref{PEP 2}, it can be seen that the exponent of the SNR term $\gamma$ is $r$, which illustrates that the rank of difference matrix $\bm{\Delta}(\mathbf{x},\hat{\mathbf{x}})$ dominates the PEP. Furthermore, the PEP with the minimum value of $r$ dominates the overall BEP as well. Finally, the diversity order of ODDM is defined as 
\begin{equation}
	\rho \triangleq \min_{\mathbf{x}\neq\hat{\mathbf{x}}} \; \text{rank}(\bm{\Delta}(\mathbf{x},\hat{\mathbf{x}})).
	\label{diversity order}
\end{equation}
As indicated by \eqref{diversity order}, the diversity order is closely related to the minimum rank of the difference matrix $\bm{\Delta}(\mathbf{x},\hat{\mathbf{x}})$. This minimum rank is determined by the delay and Doppler shift, denoted as $l_p$ and $k_p$. 
\rev{
	Consequently, the minimum asymptotic diversity order can be one for certain $\{l_p,k_p\}$, while the potential maximum diversity order can reach $P$.
	In the asymptotic region, the diversity order governs the decay rate of the BER as the SNR increases, following $\text{BER} \propto \text{SNR}^{-\rho}$.
}

\rev{Moreover, we derive a lower bound on the BER of ODDM. This lower bound, along with the simulation results, can provide the insights into the trends of the BER performance with SNR increasing.} According to \cite{2019SurabhiDiversity}, the lower bound on the BER can be obtained by summing the PEPs corresponding to all the pairs $\mathbf{x}$ and $\hat{\mathbf{x}}$, whose difference matrix $\bm{\Delta}(\mathbf{x},\hat{\mathbf{x}})$ has the rank that equals to one. With this assumption, the lower bound on the BER is given by
\begin{equation}
	\text{BER} \ge \frac{1}{Q^{MN}}\sum_{\text{rank}=1}P(\mathbf{x}\rightarrow\hat{\mathbf{x}}).
\end{equation}
When $\bm{\Delta}(\mathbf{x},\hat{\mathbf{x}})$ has rank one, the only one non-zero singular value of the difference matrix equals to $\sqrt{4PMN}$.
Therefore, the PEP with $\bm{\Delta}(\mathbf{x},\hat{\mathbf{x}})$ having rank one can be simplified to
\begin{equation}
	P(\mathbf{x}\rightarrow\hat{\mathbf{x}})=\mathbb{E}\left[\mathcal{Q}\left(\sqrt{2\gamma PMN |\tilde{h}_1|^2}\right)\right].
	\label{rank 1 PEP}
\end{equation}
Since $\tilde{h}_1 \sim \mathcal{CN}\left(0,1/P\right)$, evaluating the expectation in \eqref{rank 1 PEP} gives \cite{2005TseFundamentals}
\begin{equation}
	P\left(\mathbf{x}\rightarrow\hat{\mathbf{x}}\right)=\frac{1}{2}\left(1-\sqrt{\frac{MN}{MN+\gamma^{-1}}}\right).
\end{equation}
Therefore, the lower bound is obtained as
\begin{equation}
	\text{BER}\ge \frac{\kappa}{Q^{MN}}\frac{1}{2}\left(1-\sqrt{\frac{MN}{MN+\gamma^{-1}}}\right),
\end{equation}
where $\kappa$ denotes the number of difference matrices having rank one.
At high SNRs, this expression can be further simplified to 
\begin{equation}
	\text{BER}\ge \frac{\kappa}{Q^{MN}}\frac{1}{4\gamma MN}.
	\label{simplified lower bound}
\end{equation}
Observes that \eqref{simplified lower bound} serves as the approximated diversity one lower bound on the average BER and its value depends on the ratio $\frac{\kappa}{Q^{MN}}$. As the values $M$ and $N$ increase, the $Q^{MN}$ term dominates the ratio $\frac{\kappa}{Q^{MN}}$, hence increasing $M$ and $N$ can reduce the lower bound in \eqref{simplified lower bound}.

\begin{figure*}[t]
	\vspace*{-25pt}
	\begin{equation}
		\bm{\Xi}_p[\cdot,q]=\begin{cases}
			e^{j2\pi\frac{k_p(m-l_p)}{MN}}, \quad & \text{if } m \ge l_p \text{ and } q=\left[m-l_p\right]_M+M\left[n-k_p\right]_N \\
			e^{j2\pi\frac{k_p(m-l_p)}{MN}}e^{-j2\pi\frac{[n-k_p]_N}{N}}, \quad & \text{if } m < l_p \text{ and } q=\left[m-l_p\right]_M+M\left[n-k_p\right]_N \\
			0, \quad & \text{otherwise}
		\end{cases}
		\label{Xi definition}
	\end{equation}
	\rule{\linewidth}{.5pt}
	\vspace*{-25pt}
\end{figure*}

\section{\rev{OAMP Based Detector For ODDM}}
\rev{In this section, we briefly introduce OAMP and explore its application in ODDM modulation.}
The OMAP algorithm is designed to solve the sparse linear inverse problem like \eqref{sparse linear problem}, and it refers to the following recursion\rev{:}
\begin{subequations}
	\begin{align}
		& \text{LE: }\mathbf{r}_t=\mathbf{x}_t+\mathbf{W}_t\left(\mathbf{y}-\mathbf{Hx}_t\right) \\
		& \text{NLE: }\mathbf{x}_{t+1}=\eta_t(\mathbf{r}_t),
	\end{align}
	\label{iteration model}
\end{subequations}
where $\eta_t$ is a component-wise Lipschitz continuous function of $\mathbf{r}_t$. 
\rev{This OAMP based detector is decomposed into the iteration of a ``divergence-free'' linear estimator (LE) and a ``divergence-free'' non-linear estimator (NLE) \cite{2017OAMP}.}
The LE aims to eliminate the effects of multipath interference, then the NLE is designed to reduce the Gaussian noise and improve the quality of estimation.
Through operating the two estimators iteratively, the posterior probability $P(\mathbf{x}|\mathbf{y},\mathbf{H})$ can be decoupled into a series of $P(x_i|\mathbf{y},\mathbf{H})$.

As shown in \eqref{iteration model}, the main idea of OMAP is to design a linear transformer $\mathbf{W}_t$ and a denoiser $\eta_t\left(\cdot\right)$ satisfying orthogonality and divergence-free. 
\rev{For convenience, we define the input and output errors as $\mathbf{q}_t=\hat{\mathbf{x}}_t-\mathbf{x}$ and $\mathbf{p}_t=\mathbf{r}_t-\mathbf{x}$, where $\hat{\mathbf{x}}_t$ is the estimation of $\mathbf{x}$ at the $t$-th iteration.}
Also, we define the error variance of the LE and NLE as
\begin{equation}
	\nu_t^2=\frac{1}{MN}\mathbb{E}\{\|\mathbf{q}_t\|_2^2\}, \quad \tau_t^2=\frac{1}{MN}\mathbb{E}\{\|\mathbf{p}_t\|_2^2\}.
\end{equation}
\rev{Furthermore, the structure of $\mathbf{W}_t$ is derived from the LMMSE decorrelated matrix, which implies
	\begin{equation}
		\mathbf{W}_t=\frac{MN}{\mathrm{Tr}\left(\hat{\mathbf{W}}_t\mathbf{H}\right)}\hat{\mathbf{W}}_t,
		\label{LMMSE preprocessing matrix}
	\end{equation}
	where $\hat{\mathbf{W}}_t=\mathbf{H}^\text{H}\left(\mathbf{H}\mathbf{H}^\text{H}+\frac{\sigma^2}{\nu_t^2}\mathbf{I}\right)^{-1}$. }
This linear estimator aims to decouple the linear mixing model \eqref{sparse linear problem} into $MN$ parallel models, i.e., $r_t=x+w_t$.

Henceforth, the principles of the OAMP based detector for ODDM can be summarized in Algorithm \ref{OAMP algorithm}, where $\mathbf{B}_t=\mathbf{I}-\mathbf{W}_t\mathbf{H}$.
\begin{algorithm}
	\caption{OAMP based detector for ODDM}
	\label{OAMP algorithm}
	\begin{algorithmic}[1]
		\Statex \textbf{Input: } $\mathbf{y}$,$\mathbf{H}$, the maximum number of iterations $T$
		\Statex \textbf{Output: } $\hat{\mathbf{x}}$
		\Statex \textbf{Initialization: } $\hat{\mathbf{x}}_0=\mathbf{0}$, $\nu_0^2=1$
		\While{$t\le T$}
		\rev{
			\begin{subequations}
				\begin{align}
					&\mathbf{r}_t=\hat{\mathbf{x}}_t+\mathbf{W}_t\left(\mathbf{y}-\mathbf{H}\hat{\mathbf{x}}_t\right) \label{linear estimator} \\
					&\tau_t^2=\frac{1}{2MN}\mathrm{Tr}\left(\mathbf{B}_t\mathbf{B}_t^\text{H}\right)\nu_t^2+\frac{1}{4MN}\mathrm{Tr}\left(\mathbf{W}_t\mathbf{W}_t^\text{H}\right)\sigma^2 \label{error variance update 1} \\
					&\hat{\mathbf{x}}_{t+1}=\mathbb{E}\{\mathbf{x}|\mathbf{r}_t,\tau_t\} \label{nonlinear estimator} \\
					&\nu_{t+1}^2=\frac{\|\mathbf{y}-\mathbf{H}\hat{\mathbf{x}}_{t+1}\|_2^2-MN\sigma^2}{\mathrm{Tr}\left(\mathbf{H}^\text{H}\mathbf{H}\right)} \label{error variance update 2}
				\end{align}
			\end{subequations}
		}
		\EndWhile
	\end{algorithmic}
\end{algorithm}

Furthermore, the denoiser is given by
\begin{equation}
	\mathbb{E}\{\mathbf{x}|\mathbf{r}_t,\tau_t\}=\frac{\sum_{x_i\in\mathbb{A}}x_ie^{-\frac{\|r_i-x_i\|_2^2}{\tau_t^2}}}{\sum_{x_i\in\mathbb{A}}e^{-\frac{\|r_i-x_i\|_2^2}{\tau_t^2}}}.
\end{equation}

The Bayesian MMSE estimator and the posterior mean estimator in Algorithm \ref{OAMP algorithm} are orthogonal if the channel matrix is unitary invariant. 
By applying the element-wise estimation to the decoupled signal, the OMAP algorithm is Bayes optimal.

\rev{
	Shown by Algorithm~\ref{OAMP algorithm}, the computational complexity of the OAMP based detector mainly comes from LMMSE preprocessing matrix \eqref{LMMSE preprocessing matrix}, linear and nonlinear estimators \eqref{linear estimator} \eqref{nonlinear estimator} and residual variance update \eqref{error variance update 1} \eqref{error variance update 2}. 
	In \eqref{LMMSE preprocessing matrix}, the complexity relates to the matrix inversion with the order $\mathcal{O}\left((MN)^3\right)$. Note that the operation is performed only once throughout the algorithm.
	The complexity order of \eqref{linear estimator} \eqref{nonlinear estimator} \eqref{error variance update 1} \eqref{error variance update 2} is all $\mathcal{O}(MN)$ in each iteration due to the element-wise operation. 
	Considering the number of iterations is $T$, the overall computational complexity for the detector can be $\mathcal{O}\left((MN)^3+TMN\right)$.
}

\section{Simulation Results}

\subsection{\rev{Simulation Validation for Analysis}}
In this subsection, we perform the simulations under the simplified assumption to illustrate the upper and lower bound of the BER performance with the ML detection. For ease of simulations, the system model is simplified to $M=N=2$, and the modulation order is set to 4QAM. The channel model is set according to \eqref{H matrix}, and the number of paths is set to $P=4$. The path delay taps are chosen from $l_p\in\{0,\cdots,M-1\}$, and the Doppler taps are chosen from $k_p\in\{0,\cdots,N-1\}$. Especially, $E_\text{s}$ is the energy allocated to each symbol, and $E_\text{s}/N_0$ illustrates the transmit signal-to-noise ratio (SNR).
\rev{$T$ in the legends represents the number of iterations.}

\begin{figure}[htbp]
	\vspace*{-20pt}
	\centering
	\subfloat[Upper bound, lower bound, and the simulated BER performance.]{\label{Theoretical performance}\includegraphics[width=0.5\linewidth]{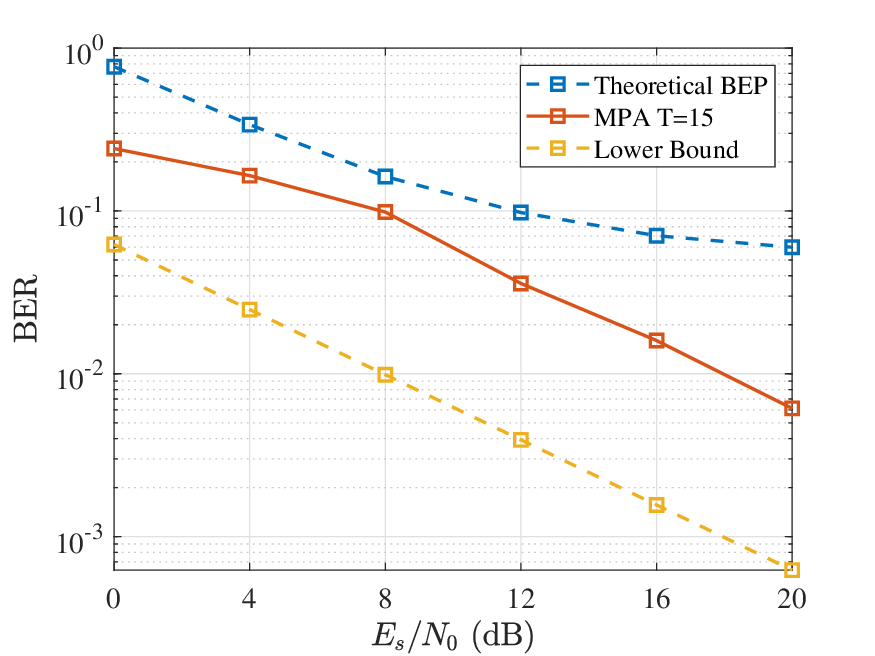}} 
	\subfloat[BER performance and lower bound for different frame sizes.]{\label{frame size}\includegraphics[width=0.5\linewidth]{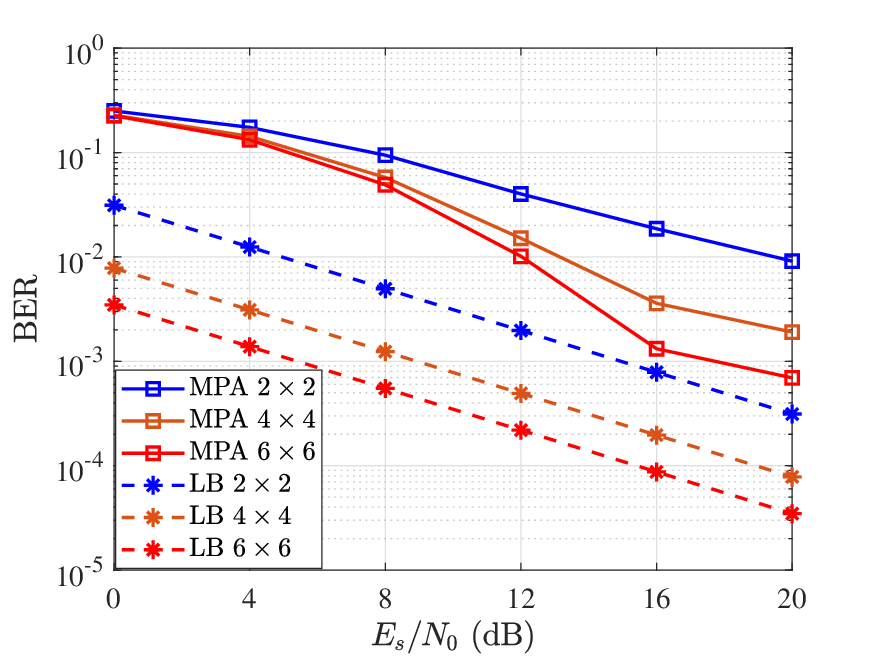}}
	\caption{The theoretical and simulated performance of ODDM versus $E_b/N_0$.}
	\vspace*{-10pt}
\end{figure}

\rev{
	In addition to the simulated BER curves, in Fig.~\ref{Theoretical performance}, we also plot the PEP as the upper bound and the lower bound from \eqref{simplified lower bound} for the considered system. It is observed that the simulated BER performance curve lies between the lower bound and the upper bound. Although the bound is not tight, it can still provide the insights for us from the trends. It can be seen from the figure that the simulated BER shows a higher diversity order in the medium to high $E_\text{s}/N_0$ region, before its slope changes to be close to that of the lower bound.
	Fig.~\ref{frame size} shows the BER performance and lower bound for three systems of different frame sizes $\{M,N\}$. It is observed that the BER curves in all three systems meet the diversity order one lower bound at the high SNR region. Meanwhile, the BER curves with the larger frame sizes show a diversity order higher than one in the low to medium SNR region before meeting the diversity order one. This means that the BER curves can decrease with a higher slope for larger $\{M,N\}$ before it changes the slope and meets the diversity order one lower bound of \eqref{simplified lower bound}.  
}

\subsection{Practical Performance Comparison}
This subsection illustrates the BER performance of the OAMP detector for ODDM under the NTN-TDL channel model given by 3GPP \cite{3GPPTR38.811}. 
The system with $M=64,N=16$ employs 4QAM symbol mapping. 
The carrier frequency of C band and the subcarrier spacing of $15\;\text{kHz}$ are considered. 
The delay spread employs $\tau_\text{s}=300\;\text{ns}$ and the Doppler spread employs $\nu_\text{s}=3750\;\text{Hz}$ for channel generation.

\begin{figure*}[tb]
	\rev{
		\centering
		\subfloat[The BER performance of ODDM with various detectors.]{\label{BER comparison B}\includegraphics[width=0.25\linewidth]{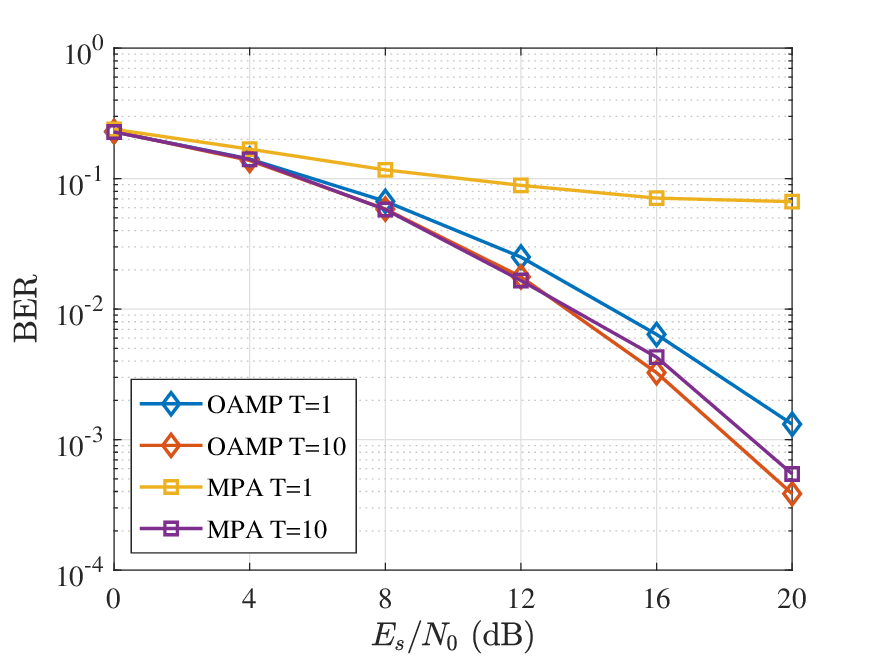}} 
		\subfloat[The iterations of various detectors with $E_\text{s}/N_0=16$ dB.]{\label{iteration comp}\includegraphics[width=0.25\linewidth]{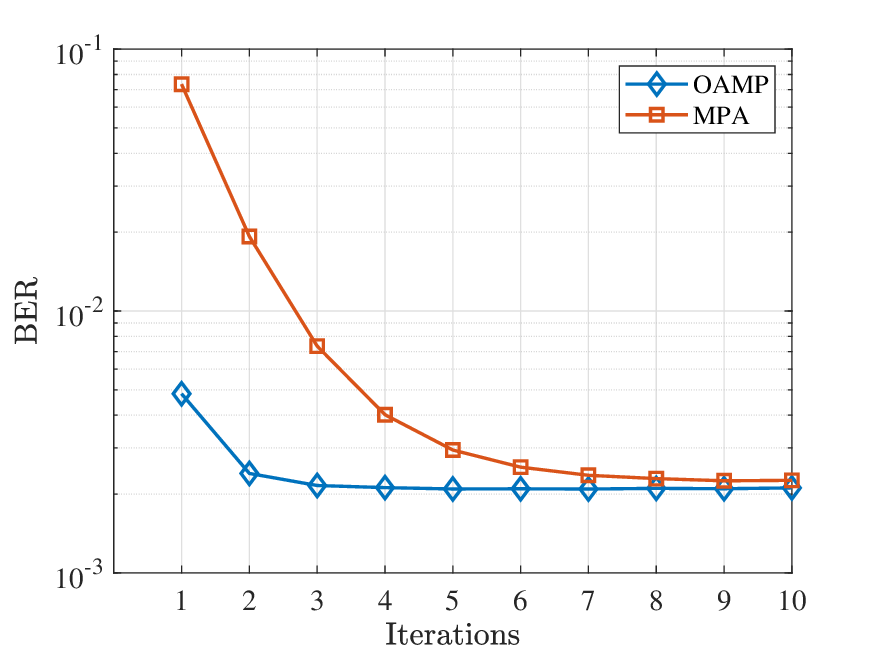}}
		\subfloat[The BER performance of various modulation methods.]{\label{modulation comparison}\includegraphics[width=0.25\linewidth]{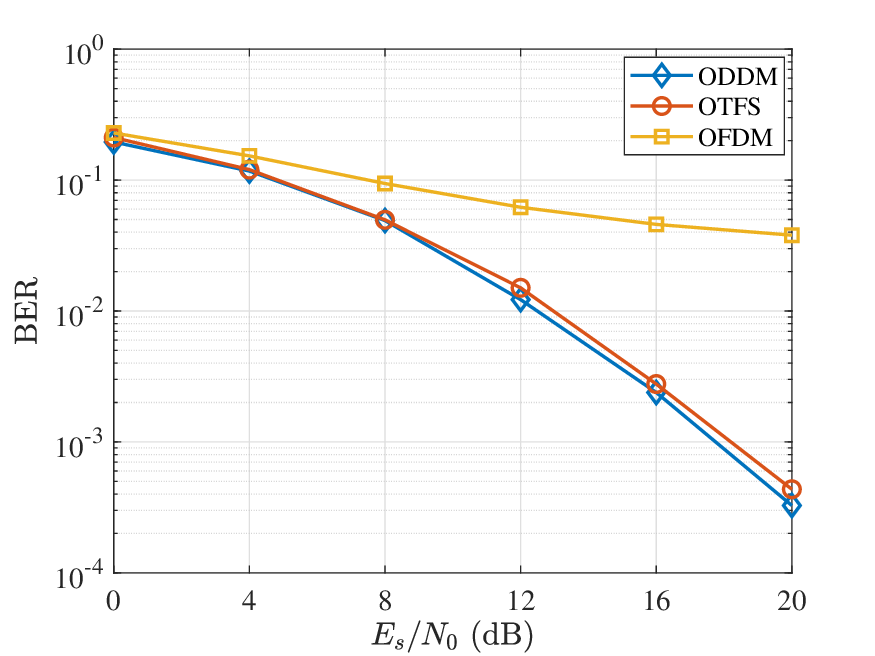}}
		\subfloat[PSD comparison, $M=512$, $N=32$, 16-QAM.]{\label{PSD comparison}\includegraphics[width=0.25\linewidth]{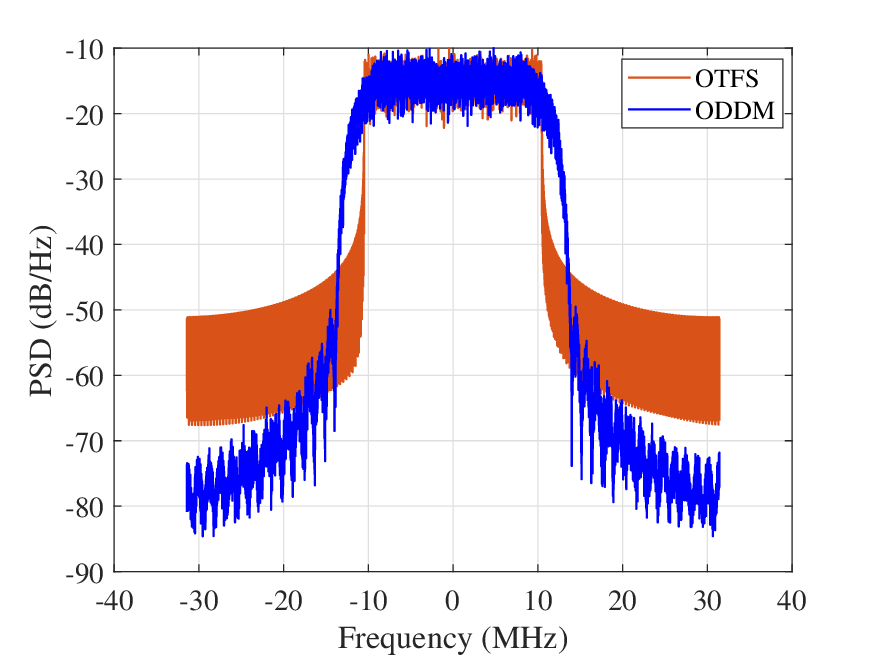}}
		\caption{\rev{The simulation results under NTN-TDL-B channel.}}
	}
\end{figure*}

\rev{In Fig.~\ref{BER comparison B}, we compare the BER performance of the OAMP based detector and the MPA based detector with different iteration settings under NTN-TDL-B channels. We observe that with $E_\text{s}/N_0$ increasing, the BER performance of both detectors becomes better.
	It can be clearly observed that the OAMP based detector performs better than the MPA based detector significantly when the number of iterations is low. With one iteration, the OAMP based detector outperforms the MPA based detector by up to $20\;\text{dB}$ in the high SNR region. }
\rev{	
	In Fig.~\ref{iteration comp}, we observe that the MPA based detector reaches its performance ceiling at approximately 10 iterations, while the OAMP based detector achieves a comparable performance with only 3 iterations.
	The comparison between two detectors with different iterations demonstrate that the OAMP based detector can achieve comparable performance to the MPA based detector with fewer iterations, highlighting its advantage in terms of reduced computational complexity.
	This phenomenon illustrates the superior efficiency of the OAMP based detector, which benefits from its enhanced  exploitation of the sparsity in the effective DD domain channel.}

\rev{In Fig.~\ref{modulation comparison}, we conduct a comparison of the BER performance of various modulation methods under the NTN-TDL-B channel.
	In this comparison, ODDM and OTFS modulation both utilize the OAMP detector with 10 iterations, while OFDM modulation employs 1tap method proposed in \cite{2018HongMP}.
	The simulations show that both OTFS and ODDM modulation outperform OFDM modulation, indicating the advantages of exploiting DD domain diversity. Moreover, the BER performance of two DD domain modulation methods is found to be quite close. 
	This phenomenon can be attributed to the fact that the asymptotic diversity order of OTFS is also one, which is derived in detail in Appendix A. The same diversity order results in the performance trend similar to that of ODDM.
}

\rev{In addition, we examine the power spectral density (PSD) of transmitted DD communication signals in Fig.~\ref{PSD comparison}. 
	Because of the square-root Nyquist pulses based shaping, ODDM exhibits significant OOBE reduction comparing to OTFS, achieving an improvement of up to 30 dB at the expense of excess bandwidth.}
Due to the severe OOBE, the bandwidth of a fully-loaded OTFS system based on the TF domain OFDM interpretation is not well-defined. 
\rev{Therefore, in order to sharpen the spectrum, the SFFT based OTFS systems can not fully utilize all subcannels as some null subcarriers placed at the band edges, leading to a reduced spectral efficiency.}
Furthermore, because the information-carrying symbols are modulated in the DD domain, it is still unclear how to arrange them such that the edge subcarriers in the TF domain remain unloaded after the ISFFT.
On the other hand, ODDM systems can fully utilize the subchannels. By adjusting the roll-off factor, a trade-off between the excess bandwidth and OOBE can be struck to achieve the desirable spectral efficiency.  
This observation verifies the practical advantage of ODDM.



\section{Conclusion}
\rev{
In this paper, we have investigated the DD diversity order of ODDM modulation and demonstrated that the minimum asymptotic DD diversity order with ML detection can be one. It is found that as frame sizes increase, the diversity order approaches one at lower BER values.
These observations have been illustrated via the BER lower bound derivation and numerical simulations.
We have presented an OAMP based detector for ODDM to extract full diversity in the DD domain with the reduced complexity.
The simulation results have demonstrated that ODDM achieved comparable performance to OTFS with the significantly reduced OOBE.
}

\appendix
\section{Diversity Analysis of ODDM}
Consider a point-to-point OTFS system, the effective DD domain input-output relation can be vectorized as 
\begin{equation}
	\mathbf{y}=\mathbf{Hx}+\mathbf{w},
	\label{OTFS relation 0}
\end{equation}
where $\mathbf{x}\in\mathbb{A}^{MN\times 1}$, $\mathbf{y,w} \in \mathbb{C}^{MN\times 1} $, $\mathbf{H}\in\mathbb{C}^{MN\times MN}$, and $\mathbf{w}\sim\mathcal{CN}\left(\mathbf{0},N_0\mathbf{I}_{MN}\right)$.
The DD domain effective channel matrix $\mathbf{H}$ is transformed from the $P$-path sparse time-delay (TD) domain channel matrix $\mathbf{H}_\text{TD} \in \mathbb{C}^{MN\times MN}$, 
\begin{equation}
	\mathbf{H}_\text{TD}=\sum_{i=1}^{P}\bm{\Pi}^{l_i}\bm{\Delta}^{k_i},
\end{equation}
with $\bm{\Pi}$ the $MN$-point permutation matrix (forward cyclic shift), $\bm{\Delta}\triangleq \text{diag}\left[z^0,z^1,\cdots,z^{MN-1}\right]$, $z=e^{\frac{j2\pi}{MN}}$. 
The effective DD domain channel matrix $\mathbf{H}$ can be obtained via
\begin{equation}
	\mathbf{H} \triangleq \left(\mathbf{F}_N\otimes\mathbf{I}_M\right)\mathbf{H}_\text{TD}\left(\mathbf{F}_N^\text{H}\otimes\mathbf{I}_M\right),
\end{equation}
where $\mathbf{F}_N$ is the $N$-point Discrete Fourier matrix. \\
Note that there are only $P$ non-zero elements in each row and column of $\mathbf{H}$, the input-out relation in \eqref{OTFS relation 0} can be rewritten in an alternate form as
\begin{equation}
	\mathbf{y}^\text{T}=\mathbf{h}^\prime\mathbf{X}+\mathbf{w}^\text{T},
\end{equation} 
where $\mathbf{h}\prime$ is a $1\times P$ vector whose $i$-th entry is given by $h_ie^{-j2\pi\tau_i\nu_i}$, and $\mathbf{X}$ is a $P\times MN$ matrix whose $i$-th column ($i=k+Nl, \; i=0,1,\cdots,MN-1$), denoted by $\mathbf{X}[i]$, is given by
\begin{equation}
	\mathbf{X}[i]=\left[ \begin{array}{c}
		x_{\left[ k-\beta _1 \right] _N+N\left[ l-\alpha _1 \right] _M}\\
		x_{\left[ k-\beta _2 \right] _N+N\left[ l-\alpha _2 \right] _M}\\
		\vdots\\
		x_{\left[ k-\beta _P \right] _N+N\left[ l-\alpha _P \right] _M}\\
	\end{array} \right] 
\end{equation}.
The representation of $\mathbf{X}$ in the alternate form allows us to view $\mathbf{X}$ as a $P\times MN$ symbol matrix. For convenience, we normalize the elements of $\mathbf{X}$ so that the average energy per symbol time is one. The SNR $\gamma$ is therefore given by $\gamma=1/N_0$. Assuming perfect channel state information and ML detection at the receiver, the PEP between $\mathbf{X}_i$ and $\mathbf{X}_j$ is given by
\begin{equation}
	P(\mathbf{X}_i\rightarrow \mathbf{X}_j|\mathbf{h}^{\prime},\mathbf{X}_i)=\mathcal{Q}\left( \sqrt{\frac{\parallel \mathbf{h}^{\prime}(\mathbf{X}_i-\mathbf{X}_j)\parallel ^2}{2N_0}} \right) .
\end{equation}
The PEP averaged over the channel statistic is given by
\begin{equation}
	P(\mathbf{X}_i\rightarrow \mathbf{X}_j)=\mathbb{E} \left[\mathcal{Q}\left( \sqrt{\frac{\gamma \parallel \mathbf{h}^{\prime}(\mathbf{X}_i-\mathbf{X}_j)\parallel ^2}{2}} \right) \right].
	\label{OTFS PEP average 0}
\end{equation}
As in ODDM, \eqref{OTFS PEP average 0} can be obtained as
\begin{equation}
	P(\mathbf{X}_i\rightarrow \mathbf{X}_j)=\mathbb{E} \left[\mathcal{Q}\left( \sqrt{\frac{\gamma \sum_{l=1}^r{\lambda _{l}^{2}|\tilde{h}_l|^2}}{2}} \right) \right],
	\label{OTFS PEP average2 0}
\end{equation}
where $r$ denotes the rank of the difference matrix $\bm{\Delta}_{ij}=\left(\mathbf{X}_i-\mathbf{X}_j\right)$, $\tilde{h}_l$ is the $l$-th element of the vector $\tilde{\mathbf{h}}^\text{H}=\mathbf{U}^\text{H}{\mathbf{h}^\prime}^\text{H}$, where the matrix $\left(\mathbf{X}_i-\mathbf{X}_j\right)\left(\mathbf{X}_i-\mathbf{X}_j\right)^\text{H}$ is the Hermitian matrix that is diagonalizable by unitary transformation. Hence, it can be written as $\left(\mathbf{X}_i-\mathbf{X}_j\right)\left(\mathbf{X}_i-\mathbf{X}_j\right)^\text{H}=\mathbf{U}\bm{\Lambda}\mathbf{U}^\text{H}$, where $\mathbf{U}$ is unitary and $\bm{\Lambda}=\text{diag}\{\lambda_1^2,\lambda_2^2,\cdots,\lambda_P^2\}$, $\lambda_i$ being the $i$-th singular value of the difference matrix $\bm{\Delta}_{ij}$.
The average PEP in \eqref{OTFS PEP average2 0} can be simplified to get the following upper bound,
\begin{equation}
	P(\mathbf{X}_i\rightarrow \mathbf{X}_j)\le \prod_{l=1}^r{\frac{1}{1+\,\,\frac{\gamma \lambda _{l}^{2}}{4P}}}.
\end{equation}
At high SNRs, it can be further simplified as
\begin{equation}
	P(\mathbf{X}_i\rightarrow \mathbf{X}_j)\le \frac{1}{\gamma ^r\prod_{l=1}^r{\frac{\lambda _{l}^{2}}{4P}}}.
	\label{OTFS simplified PEP 0}
\end{equation}
From \eqref{OTFS simplified PEP 0}, it can be seen that the exponent of the SNR term $\gamma$ is $r$, which is equal to the rank of the difference matrix $\bm{\Delta}_{ij}$. For all $i\neq j$, the PEP with the minimum value of $r$ dominates the overall BER. Therefore, the achieved diversity order $\rho$ is given by
\begin{equation}
	\rho=\min_{i\neq j} \, \text{rank}(\bm{\Delta}_{ij}).
\end{equation}
Now ,consider a case where $\mathbf{X}_i=a.\mathbf{1}_{P\times MN}$ and $\mathbf{X}_j=a^\prime.\mathbf{1}_{P\times MN}$. Then, $\bm{\Delta}_{ij}=(\mathbf{X}_i-\mathbf{X}_j)=(a-a^\prime).\mathbf{1}_{P\times MN}$, whose rank is one, indicates the minimum rank of $\bm{\Delta}_{ij},\,\forall i,j, \, i\neq j$.
Hence, the minimum asymptotic diversity order of OTFS with ML detection is one.

\bibliographystyle{IEEEtran}
\bibliography{myre}

\end{document}